\documentclass[prl, twocolumn, floatfix]{revtex4}
\usepackage{graphicx}
\usepackage{color}
\usepackage{amsmath, amsfonts, amssymb, bm}
\usepackage{dsfont,slashed}
\usepackage{nicefrac}

\DeclareFontFamily{OT1}{pzc}{}
\DeclareFontShape{OT1}{pzc}{m}{it}{<-> s * [1.10] pzcmi7t}{}
\DeclareMathAlphabet{\mathpzc}{OT1}{pzc}{m}{it}

\begin{document}
\title{Dimensionality-driven photoproduction of massive Dirac pairs near threshold \\ in gapped graphene monolayers}
\author{A. Golub}
\author{R. Egger}
\author{C. M\"uller}
\author{S. Villalba-Ch\'avez}
\affiliation{Institut f\"ur Theoretische Physik, Heinrich Heine Universit\"at D\"usseldorf, Universit\"atsstr. 1, 40225 D\"usseldorf, Germany}
\date{\today}
\begin{abstract}
Generation of quasi-particle--hole pairs in gapped graphene monolayers in the combined field of two counterpropagating light waves is studied. The process represents an analogue of electron-positron pair production from the QED vacuum by the Breit-Wheeler effect. We show, however, that the two-dimensional structure of graphene causes some striking differences between both scenarios. In particular, contrary to the QED case, it allows for non-zero pair production rates at the energy threshold when the Breit-Wheeler reaction proceeds nonlinearly with absorption of three photons.
\end{abstract}
\maketitle

{\it Introduction.}---When atoms or molecules interact with strong laser fields, nonlinear processes relying on the absorption of multiple photons can occur due to the very high photon densities in the field. A variety of strong-field phenomena was discovered, ranging from multiphoton ionization to high-harmonic generation \cite{Review1}, which have paved the way towards new research areas \cite{atto}.

In recent years there is a growing interest in interactions of intense laser fields with condensed matter systems \cite{Review_solids}. While some of the basic strong-field concepts established in atomic physics can be 
applied to solids as well \cite{solids1,solids2}, their more complex electronic structure renders laser-solid interactions generally more involved. In particular, condensed-matter systems can be distinguished by the 
geometric arrangement of atoms, the symmetry and topology of the band structure, and the behavior of electrons therein. The question thus arises which signatures may emerge in intense laser-solid interactions from the 
characteristic properties of the system. For example, distinct topological effects from edge states in linear chains have recently been predicted for the strong-field process of high-harmonic generation in solids \cite{Bauer}. 

A solid-state system of special relevance for strong-field studies is graphene \cite{graphene} because it can withstand high laser intensities. It is distinguished by its two-dimensional geometry, forming a monolayer of 
carbon atoms, and its peculiar electronic properties. In a vicinity of the Fermi surface, the electrons exhibit a dispersion relation like Dirac fermions, with the speed of light $c$ replaced by the Fermi velocity 
$v_F\approx c/300$. Recent experiments have demonstrated coherent control of electron dynamics in graphene by driving Landau-Zener transitions with phase-stabilized short laser pulses \cite{Hommelhoff}, including the passage 
from the weak-field to the strong-field regime \cite{Hommelhoff_NJP}. Field-driven acceleration of Dirac fermions, which also exist in topological insulators \cite{Review_top}, has been observed by combined irradiation 
of bismuth-telluride surfaces with intense tera\-hertz and short ultraviolet pulses \cite{surface}.

Graphene moreover serves as test ground for strong-field processes from the realm of quantum electrodynamics (QED). Fundamental phenomena, such as Klein tunneling \cite{Klein}, Casimir force \cite{Casimir} or Coulomb 
supercriticality \cite{supercriticality}, find their low-energy counterpart in graphene. Theoreticians have also discovered graphene as a means to study the Schwinger effect, i.e. the spontaneous production of electron-positron 
pairs in a constant electric field $E_0$ \cite{McGady,Lewkowicz, Moessner, Avetissian, Mostepanenko, Fillion-Gourdeau, Akal2016, Akal2019}. The similarity with QED is particularly close in bandgap graphene where the 
quasiparticles acquire a nonzero mass \cite{bandgap}. The Schwinger rate has the characteristic form $\mathcal{R}_{\rm S}\sim E_0^\nu\exp(-\pi E_c/E_0)$ with the critical field strength $E_c$ associated with the particle 
mass. The different geometry of the underlying vacuum state only exerts a minor impact here by changing the power of the pre-exponential factor from $\nu=2$ in QED to $\nu=3/2$ in graphene \cite{McGady, Akal2016}.

\begin{figure}[b]  
\vspace{-0.25cm}
\begin{center}
\includegraphics[width=0.48\textwidth]{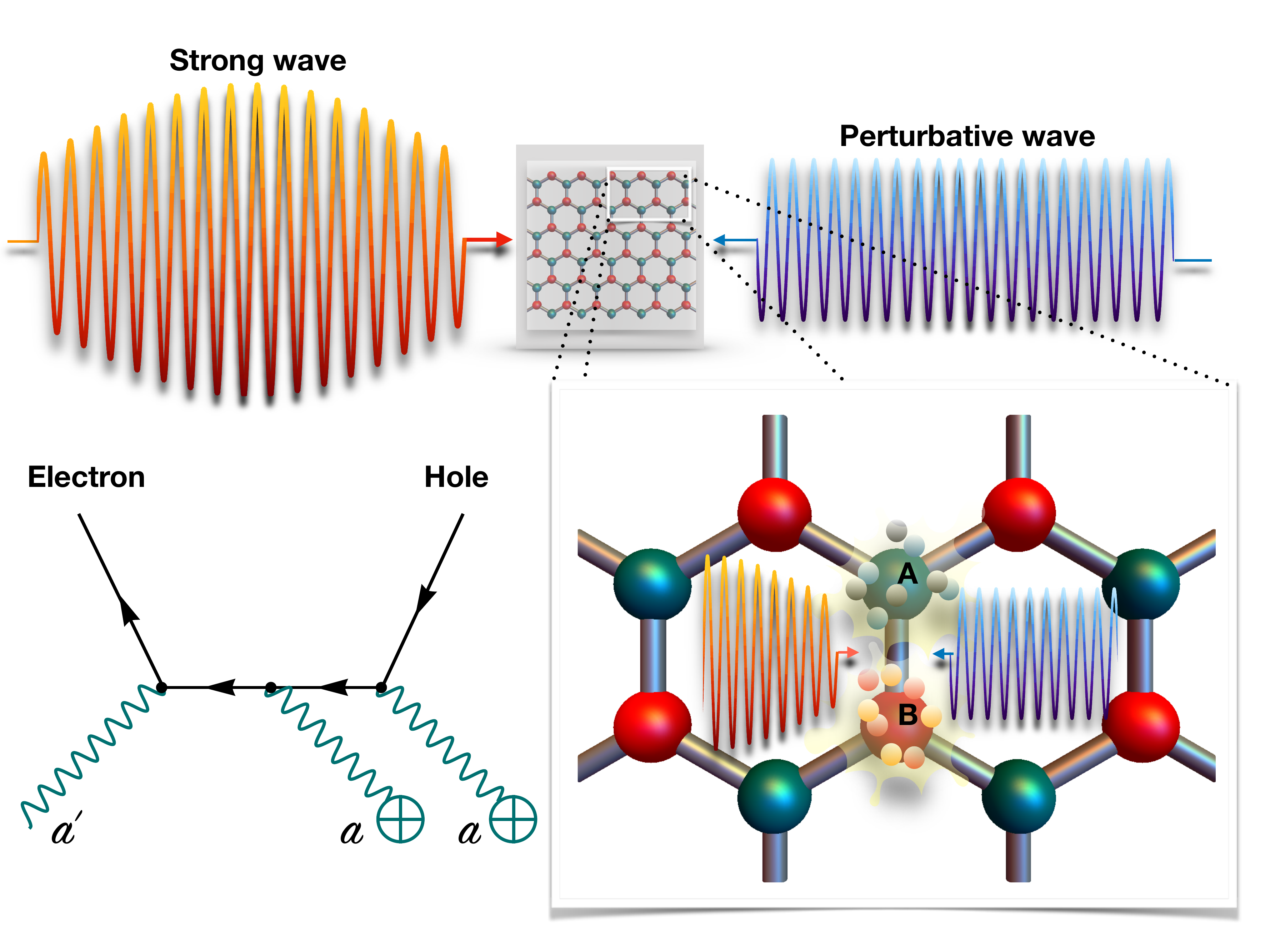}
\end{center}
\vspace{-0.5cm} 
\caption{Scheme of Breit-Wheeler-type production of massive Dirac pairs in a gapped graphene monolayer. The enlarged section illustrates graphene's honeycomb lattice composed of two sublattices $A$ and $B$. The reciprocal 
lattice in momentum space is hexagonal, as well, with inequivalent $\pmb{K}$ and $\pmb{K^\prime}$ points  in the corners \cite{graphene}. A representative  Feynman diagram is shown which contributes to the three-photon reaction.}
\label{scheme}
\end{figure}

In this Letter, we study another strong-field process in graphene which has its counterpart in QED. When a graphene sheet is exposed to the electromagnetic field of two counterpropagating light waves, quasi-particle--hole 
pairs can be generated by photon absorption (see Fig.~\ref{scheme}). This process represents the analog of (non)linear Breit-Wheeler [(N)LBW] pair production from the QED vacuum \cite{BW,Ritus,other,SLAC}. It may be written symbolically as
\begin{equation}
n k + k' \to p_- + p_+\ ,
\label{NLBW}
\end{equation}
where $n$ denotes the number of photons with wave vector $k$ from a strong laser field and $k'$ is the wave vector of a weak counterpropagating wave, whereas $p_{\pm}$ stand for the momenta of created electron and hole, respectively. 
The NLBW process of QED was observed in an intermediate coupling regime ($n\sim 5$) in collisions of a highly relativistic electron beam with an intense optical laser pulse \cite{SLAC}. The original (i.e. linear) Breit-Wheeler process with $n=1$ \cite{BW}, however, has not been measured yet. Corresponding theoretical proposals to facilitate its detection have been made in recent years which rely on various kinds of gamma-ray sources in the MeV-GeV energy range \cite{Pike,Lobet,Drebot,Yu}.
Here we provide a theoretical description of the (N)LBW process in gapped graphene monolayers and show that, in principle, they can offer a low-energy alternative, similarly to the Schwinger effect \cite{McGady,Lewkowicz, Moessner, Avetissian, Mostepanenko, Fillion-Gourdeau, Akal2016, Akal2019}. However, our analysis also reveals pronounced qualitative differences with the QED case, which emerge from the two-dimensional structure of graphene.

Planck's constant and the vacuum permittivity are set to unity, $\hbar =\epsilon_0= 1$, to simplify notation. The speed of light in vacuum is denoted by $c$ and the electron charge and mass by $e$ and $m_e$, respectively.

{\it Theoretical considerations.}---An effective field theoretical framework -- that suitably incorporates the electron-hole symmetry predicted within the nearest-neighbours tight-binding model \cite{Wallace} -- is adopted. It relies on a Lagrangian density in which the interplay between  massive Dirac quasiparticles and an electromagnetic potential $\mathcal{A}_\mu(x)$ [$\mathcal{A}_0(x)=0$], with $\mu=0,1,2$ and $x=(ct,x,y)$ occurs via a minimal coupling \cite{Gusynin, interaction}: 
\begin{equation}\label{intlagrangian}
\mathcal{L}=\sum_{\sigma=\pm 1} \bar{\Psi}_\sigma\left[ i\tilde{\gamma}^0\partial_t + v_F\tilde{\gamma}^j\left(i\partial_j - \frac{e}{c} \mathcal{A}_j\right) -\Delta\right]\Psi_\sigma.
\end{equation}
Here, $\bar{\Psi}_\sigma = \Psi^{\dagger}_\sigma\tilde{\gamma}^0$, $\Psi^T_\sigma=(\psi^T_{\sigma \pmb{K}},\psi^T_{\sigma \pmb{K}^\prime})$ is a 4-component spinor consisting of two 2-component irreducible pieces corresponding to the $\pmb{K}$ and $\pmb{K}^\prime$ points of the Fermi surface. The spinors $\psi^T_{\sigma \pmb{K}} = (\psi_{\sigma \pmb{K} A},\psi_{\sigma \pmb{K} B})$, $\psi^T_{\sigma \pmb{K}^{\prime}} = (\psi_{\sigma \pmb{K}^{\prime} B},\psi_{\sigma \pmb{K}^{\prime} A})$ combine Bloch states associated with the two different sublattices in graphene linked to atoms $A$ and $B$, whereas the electron spin is included via an additional particle flavour $\sigma$. The gamma matrices $\tilde{\gamma}^\mu$ form a reducible $4\times4$ representation, satisfying $\{\tilde{\gamma}^\mu, \tilde{\gamma}^\nu\}=2g^{\mu \nu}\mathds{1}_{4\times4}$ with the metric tensor $g^{\mu \nu}=\mathrm{diag}(1,-1,-1)$. Explicitly, we take $\tilde{\gamma}^\mu=\tau\otimes(\gamma^1,\gamma^2,\gamma^3)$ with $\tau = \sigma_3$, $\gamma^\mu = (\sigma_3, i\sigma_2,-i\sigma_1)$, and the Pauli matrices $\sigma_\ell$, $\ell=1,2,3$. The matrices $\tau$ and $\gamma^{\mu}$ act in the spaces of $\pmb{K}$, $\pmb{K}'$ points and $A$, $B$ sublattices, respectively. Moreover, in Eq.~\eqref{intlagrangian} a sum over the repeated index $j\in\{1,2\}$ is implied and we introduce a half bandgap $\Delta = m_g v_F^2$, which leads to a relativistic-like dispersion relation for quasiparticles $\varepsilon_{\pmb{p}}=\sqrt{v_F^2\pmb{p}^{2} + \Delta^2}$ relative to the Fermi-level \cite{Akal2019}. Likewise, the  momentum $\pmb{p}=(p_x,p_y)$ has to be understood relative to the $\pmb{K}$ and $\pmb{K}^\prime$ points, satisfying the condition $\vert\pmb{p}\vert\ll p_{\rm max} \approx \vert\pmb{K}\vert,\vert\pmb{K}^\prime \vert \approx 3 \ \mathrm{eV}/v_F$. 

In the following, the external field $\mathcal{A}_\mu(x) = a_\mu(x) + a_\mu^\prime(x)$ is composed of two counterpropagating electromagnetic plane waves $a_\mu(x)= \mathpzc{a}_0\epsilon_\mu \mathrm{cos}(kx)$  and 
$a'_\mu(x) = \mathpzc{a}_0^\prime\epsilon'_\mu \mathrm{cos}(k'x)$, with wave vectors $k^\mu = (\omega/c,\pmb{k})$, $k'^\mu = (\omega'/c,\pmb{k}')$ satisfying the transversality relations $\pmb{k} \cdot \pmb{\epsilon}=\pmb{k}^\prime \cdot \pmb{\epsilon}^\prime=0$. The amplitudes are supposed to fulfill the condition $\eta_g\gg  \eta_g^\prime$, in terms of the graphene-modified intensity parameters $\eta_g^{(\prime)}=\vert e\vert  \mathpzc{a}_0^{(\prime)}/(m_g v_F c)$. We will restrict ourselves to the situation in which both the propagation and polarization directions of the electromagnetic waves lie in the plane of graphene (see Fig.~\ref{scheme}) \cite{B-field}.

Being interested in the intensity regime where $\eta_g,\;\eta^\prime_g\ll 1$, we calculate the probability for (N)LBW pair production in graphene \eqref{NLBW} by applying perturbation theory in both fields. Observe that, as $ \eta_g\gg  \eta_g^\prime$, the effect of the strong plane wave can be investigated  via  coherent states, which is equivalent to considering the corresponding field as a classical background \cite{Fradkin}. Hence, the corresponding scattering operator reads
\begin{equation}
\begin{split}
\mathcal{S}
&=\hat{T}\exp\Big[ie\frac{v_F}{c}\sum_{\sigma=\pm 1}\int dt d^2 x\;\bar{\psi}_{\sigma\pmb{K}}\ \pmb{\gamma}\!\cdot\!(\pmb{a}+\pmb{a}^\prime)\psi_{\sigma\pmb{K}} \\
&\qquad \qquad \qquad \qquad \qquad +\bar{\psi}_{\sigma \pmb{K}^\prime}\ \pmb{\gamma}\!\cdot\!(\pmb{a}+\pmb{a}^\prime)\psi_{\sigma \pmb{K}^\prime}\Big]
\end{split}
\label{S}
\end{equation}
with $\bar{\psi}_{\sigma\pmb{K},\pmb{K}^\prime}=\psi^{\dagger}_{\sigma\pmb{K},\pmb{K}^\prime}\gamma^0$. Here, the field $a_\mu^\prime(x)$ is quantized and $\hat{T}$ refers, as usual, to the time-ordering  operator. The  quantization of the Dirac-like fields $\psi_{\sigma \pmb{K}}$ and $\psi_{\sigma \pmb{K}\prime}$ requires to introduce equal-time anticommutation relations, which preserve the corresponding spin and valley quantum numbers. Accordingly, we can restrict ourselves to $S$-matrix elements generated by the first term in \eqref{S} and, at the end, multiply the outcome by the spin-valley degeneracy $N_f=4$.

We remark that the treatment used here differs from the traditional path of analysing the NLBW process in QED, where the  Furry picture, based on exact solutions of the Dirac equation in the presence of a single plane-wave laser field (Volkov states), is applied \cite{Ritus,other}. The corresponding Dirac-like equation in graphene, however, is not solved by Volkov states \cite{Varro} due to the asymmetry introduced by the Fermi velocity $v_F$.

Next, we proceed to determine the $S$-matrix element describing the creation of a quasi-particle with energy-momentum $(\varepsilon_{\pmb{p}_-},\pmb{p}_-)$ and its antiparticle with $(\varepsilon_{\pmb{p}_+},\pmb{p}_+)$ by the collision of 
two photons $k^\mu$ and $k'^\mu$. In graphene this amplitude reads
\begin{eqnarray}
\label{S_1}
\begin{split}
&\langle \pmb{p}_+ \pmb{p}_-\vert\mathcal{S}_1\vert \pmb{k}^\prime\rangle = -(2\pi)^3 N_+N_-\frac{e^2 \mathpzc{a}_0 \mathpzc{a}_0^\prime v_F^2}{4c^2}\,\mathcal{M}_1\\
&\qquad\quad\times \delta(\varepsilon_{\pmb{p}_+}\!\!+\!\varepsilon_{\pmb{p}_-}\!\!-\omega-\omega')\delta^{(2)}(\pmb{p}_+\!\!+\!\pmb{p}_-\!\!-\pmb{k}-\pmb{k}'). 
\end{split}
\end{eqnarray}
Here, the $\delta$-functions encode energy-momentum conservation and $N_{\pm}=[\Delta/(\varepsilon_{\pmb{p}_{\pm}}A)]^{\nicefrac{1}{2}}$ are the normalization constants of the quantized Dirac-like field with $A$ referring to the normalization  area. Besides, the subscript $1$ in the scattering operator and the spinor-matrix product $\mathcal{M}_1$ refers to the number of photons absorbed from the classical source. The latter is given by
\begin{eqnarray}
\begin{split}
&\mathcal{M}_1=\bar{u}_{\pmb{p}_-}\left[\slashed{\epsilon}' S_g(\omega-\varepsilon_{\pmb{p}_+},\pmb{k}-\pmb{p}_+)\slashed{\epsilon}\right. \\
&\qquad\qquad\qquad\left.+\slashed{\epsilon} S_g(\omega'-\varepsilon_{\pmb{p}_+},\pmb{k}'-\pmb{p}_+)\slashed{\epsilon}' \right]v_{\pmb{p}_+}.
\end{split}
\label{M_1}
\end{eqnarray}
We point out that, in this equation the usual slash notation for products 
with $\gamma$ matrices has been employed. While $u_{\pmb{p}_-}$  and $v_{\pmb{p}_+}$ are  two-dimensional spinors fulfilling the relations $u_{\pmb{p}_{-}}\bar{u}_{\pmb{p}_{-}}=(\gamma^0\varepsilon_{\pmb{p}_-}-v_F\pmb{\gamma}\cdot\pmb{p}_- +\Delta)/(2\Delta)$, $v_{\pmb{p}_{+}}\bar{v}_{\pmb{p}_{+}}=(\gamma^0\varepsilon_{\pmb{p}_+}-v_F\pmb{\gamma}\cdot\pmb{p}_+ -\Delta)/(2\Delta)$, $S_g$  refers to the free propagator linked to  the Dirac-like field: 
\begin{equation}
S_g(\varepsilon,\pmb{p}) = \frac{i}{{\gamma^0\varepsilon-v_F\pmb{\gamma}\cdot\pmb{p}-\Delta+i0^+}}.
\end{equation}

The rate of pair production per unit area is obtained by taking the square of the $S$-matrix element, dividing through the interaction time $T_0$ and normalization area $A$, and integrating over the final density of states:
\begin{equation}
\mathcal{R}_1^{(2+1)}= N_f \int \frac{A\,d^2p_+}{(2\pi)^2} \int \frac{A\,d^2p_-}{(2\pi)^2} \frac{\vert\langle \pmb{p}_+ \pmb{p}_-\vert\mathcal{S}_1\vert \pmb{k}^\prime\rangle\vert^2}{T_0 A}\ .
\label{R_1}
\end{equation}
The field frequencies are assumed to be chosen such that the integrals are restricted to regions where $\vert\pmb{p}_\pm\vert\ll p_{\rm max}$. Equation~\eqref{R_1} contains the term $\vert\mathcal{M}_1\vert^2$ which can be rewritten as a trace over $2\times2$-gamma matrices. We point out that, due to their reduced dimensionality, several relations -- which are needed to evaluate the trace -- differ from the familiar QED$_{3+1}$ case. For instance, while in $3+1$-dimensions $\mathrm{Tr}[\gamma^\mu\gamma^\nu]=4g^{\mu \nu}$ and $\mathrm{Tr}[\gamma^\mu\gamma^\nu \gamma^\rho]=0$, in the case under
consideration one has $\mathrm{Tr}[\gamma^\mu\gamma^\nu]=2g^{\mu \nu}$ and $\mathrm{Tr}[\gamma^\mu\gamma^\nu \gamma^\rho]=-2i\epsilon^{\mu \nu \rho}$. Having these properties in mind and the fact that $ v_F/c\ll1$,
we arrive at 
\begin{eqnarray}\label{2phorateb1}
\mathcal{R}_1^{(2+1)}\approx \eta_g^2 \eta_g^{\prime 2}m_g^3 v_F^4\frac{r^2(4+\mathpzc{r}^2)}{4(1+\mathpzc{r}^2)^{\nicefrac{5}{2}}},
\end{eqnarray}
where $\mathpzc{r}=\vert\pmb{p}\vert/(m_gv_F)$ denotes a dimensionless parameter and $\omega=\omega'$ is assumed, such that 
$\pmb{p}\equiv \pmb{p}_+ = -\pmb{p}_-$ \cite{balance}.

A particularly interesting outcome results from the amplitude describing the production of pairs driven by the absorption of two classical photons and a quantized field (see Feynman diagram in Fig.~\ref{scheme}):
\begin{eqnarray}
\begin{split}
&\langle \pmb{p}_+ \pmb{p}_-\vert\mathcal{S}_2\vert \pmb{k}^\prime\rangle =-i(2\pi)^3 N_+N_-\frac{e^3 \mathpzc{a}_0^2 \mathpzc{a}_0' v_F^3}{8c^3}\,\mathcal{M}_2\\
&\qquad\quad\times \delta(\varepsilon_{\pmb{p}_+}\!\!+\varepsilon_{\pmb{p}_-}\!\!-2\omega-\omega')\delta^{(2)}(\pmb{p}_+\!\!+\!\pmb{p}_-\!\!-2\pmb{k}-\pmb{k}').
\end{split}
\end{eqnarray}
The explicit expression of  $\mathcal{M}_2$ is lengthy, and, instead of giving it explicitly, we proceed to determine the corresponding production rate $\mathcal{R}^{(2+1)}_2$. Considering the smallness of the parameter $v_F/c\ll 1$ and assuming $\omega'=2\omega$, we find 
\begin{equation} 
\mathcal{R}^{(2+1)}_2\approx\eta_g^4 \eta_g^{\prime2} m_g^3v_F^4\frac{2(4-30\mathpzc{r}^2+108 \mathpzc{r}^4+17 \mathpzc{r}^6)}{81(1+\mathpzc{r}^2)^{\nicefrac{9}{2}}}.
\end{equation}

\begin{figure}[b]  
\vspace{-0.25cm}
\begin{center}
\includegraphics[width=0.48\textwidth]{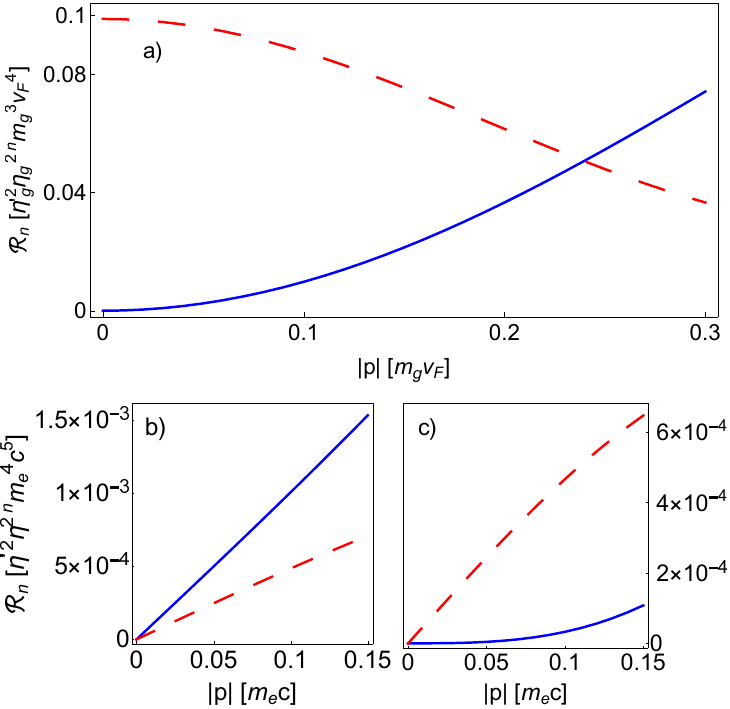}
\end{center}
\vspace{-0.5cm} 
\caption{Near-threshold behavior of the total rates for (N)LBW pair production in the center-of-mass frame, as function of the absolute value of momentum of one of the created particles: (a) in graphene, (b) in QED$_{3+1}$ with 
unpolarized $\gamma$-photon, (c) in QED$_{3+1}$ with polarized $\gamma$-photon, for $n=1$ (blue solid), $n=2$ (red dashed).}
\label{rates}
\end{figure}

{\it Results and Discussion}---Figure~\ref{rates}\,a) shows the rates $\mathcal{R}_n^{(2+1)}$ for Breit-Wheeler pair production in graphene near the energy threshold for $n\in\{1,2\}$. When $n=1$, the rate starts from zero, but for $n=2$ it attains a non-zero value at the threshold [$\mathcal{R}^{(2+1)}_2\approx 8\eta_g^4 \eta_g^{\prime 2}m_g^3v_F^4/81$]. This is in sharp contrast to the rates $\mathcal{R}_n^{(3+1)}$ for the corresponding processes in QED$_{3+1}$ which always vanish at the threshold \cite{corrections}. Thus, the special geometry of graphene leads to distinct qualitative changes in the properties of NLBW pair production.

The QED$_{3+1}$ rates scale like $\mathcal{R}_n^{(3+1)}\sim \alpha\eta^{2n}\vert\pmb{p}\vert$ for $\vert\pmb{p}\vert\to 0$, with $\eta = \vert e\vert \mathpzc{a}_0/(m_e c^2)\ll 1$, when the $\gamma$-beam is unpolarized [see Fig.~\ref{rates}\,b)]. This is in accordance with Wigner's well-known theory for the threshold behavior of quantum mechanical scattering processes with two particles in the final state \cite{threshold1}. It arises from an expression of the form 
$\int d^3p_+\int d^3p_- \vert\mathcal{M}\vert^2\,\delta^{(4)}(p_+ + p_- -Q) \propto \int dp_+^0 \vert\pmb{p}_+\vert\vert\mathcal{M}\vert^2\,\delta(p_+^0 + p_-^0 -Q^0) \propto \vert\pmb{p}_+\vert$ under the assumption that 
the squared matrix element $\vert\mathcal{M}\vert^2$ behaves like a constant near the threshold.

In the case of graphene, this argument needs to be modified. Due to the reduced dimensionality of the phase space, it reads $\int d^2p_+\int d^2p_- \vert\mathcal{M}\vert^2\,\delta^{(3)}(p_+ + p_- -Q) \propto 
\int dp_+^0 \vert\mathcal{M}\vert^2\,\delta(p_+^0 + p_-^0 -Q^0) \propto {\rm const}$. 
This explains why the NLBW pair production rate at the threshold can be non-zero in graphene, which occurs when $n=2$ [see Fig.~\ref{rates}\,a)]. A similar phenomenon is known from electron-atom scattering processes 
occuring in the presence of a strong magnetic field. The field leads to a reduction of the effective dimensionality of the problem and, thus, to a modification of its threshold behavior \cite{threshold2}.

Still, the behavior of $\mathcal{R}_n^{(2+1)}$ for $n=1$ does not follow the same law. Instead one finds $\mathcal{R}_1^{(2+1)}\sim \vert\pmb{p}\vert^2$ [see Eq.~\eqref{2phorateb1}]. Here, a more complex  explanation is required which combines the two-dimensional geometry of graphene with the polarization state of the $\omega'$-photon and the fermionic 
nature of the charge carriers. The pair production rates in graphene rely on incident light waves whose polarization vectors lie in the graphene plane and are parallel to each other. The QED$_{3+1}$ rates resulting from 
this field configuration are shown in Fig.~\ref{rates}\,c). While $\mathcal{R}_n^{(3+1)}$ for $n=2$ still grows linearly with $\vert\pmb{p}\vert$ near the threshold, a cubic dependence $\sim \vert\pmb{p}\vert^3$ is 
found for $n=1$. The fact that the produced particles are fermions is of crucial importance here. In fact, for scalar particles the rate of the ordinary Breit-Wheeler process is given by $\mathcal{R}_{1,{\rm scal}}^{(3+1)} 
\sim \alpha\eta^2 \vert\pmb{p}\vert\, (\pmb{\epsilon}\cdot\pmb{\epsilon}')^2$ \cite{Akhiezer}, which yields a linear dependence on $\vert\pmb{p}\vert$ when the photon polarizations are parallel. Instead, the leading-order 
term for production of Dirac pairs is $\mathcal{R}_1^{(3+1)} \sim \alpha \eta^2 \vert\pmb{p}\vert\, (\pmb{\epsilon}\times\pmb{\epsilon}')^2$ which vanishes in case of parallel polarization vectors. The next-to-leading order 
term grows with $\vert\pmb{p}\vert^3$, in accordance with Fig.~\ref{rates}\,c). Applying the above argument of reduced phase space to this scaling law gives the threshold behavior $\mathcal{R}_1^{(2+1)}\sim \vert\pmb{p}\vert^2$ 
found in graphene.

An experimental test of our predictions could apply moderately intense beams of terahertz radiation \cite{surface,terahertz}. If we assume, for example, a gap parameter $\Delta = 0.1$\,eV \cite{bandgap}, frequencies $\omega' = 2\omega \gtrsim 0.1$\,eV and intensities $I_0 = 10 I_0' = 10^5$\,W/cm$^2$ ($\eta_g\approx 0.1$, $\eta'_g\approx 1.6 \times 10^{-2}$), we obtain at the threshold $\mathcal{R}_2^{(2+1)} \approx  10^{10}$ s$^{-1}\mu$m$^{-2}$. We remark that, in order to be consistent with our perturbative treatment, the fields should lie well below the critical field in graphene $E_c = \Delta^2/|e|v_F$ for the considered gap. This value corresponds to the critical intensity $I_c=cE_{\mathrm{crit}}^2/2 \approx 6\times 10^7$\,W/cm$^2$. Detection of the produced pairs could be achieved by measuring the induced current when an external voltage is applied \cite{McGady}.

{\it Conclusion and Outlook}---Generation of quasi-particle--hole pairs in gapped graphene monolayers by counterpropagating photon beams was studied. The process is analogous to (N)LBW pair production in QED. Focussing on the low-intensity 
regime at moderate coupling strengths ($\eta_g,\; \eta_g' \ll 1$), we revealed striking qualitative differences between both phenomena which are caused by the different dimensionalities of the underlying vacuum state. While the pair 
production rate at the energy threshold vanishes in QED$_{3+1}$ for any photon number, in graphene it not always does. 
This result has been shown explicitly for a three-photon reaction and we expect that it holds generally when the total number of absorbed photons is odd. Corresponding even-odd staggerings appear in various strong-field processes, such as high-harmonic generation, where they are related to the total parity of photons involved. This point will be examined in a forthcoming study. 

Future work could, moreover, account for further aspects in the structure of graphene (such as edge states \cite{Bauer}) and its interaction with the external field (such as influences of the substrate on the effective field 
strength experienced by the charge carriers \cite{Hommelhoff}). However, since the modified threshold behavior results from a basic property of graphene, namely its two-dimensional geometry, the predicted effects are expected 
to be robust and to persist in improved treatments. Our results furthermore suggest that the (N)LBW process in topological matter with Dirac-like states might be used to identify domains where quasiparticle-hole generation 
is restricted to two dimensions (like, e.g., on surfaces).

{\it Acknowledgement}---This work has been funded by the Deutsche Forschungsgemeinschaft (DFG, German Research Foundation) under Grant No. 388720772 (MU 3149/5-1). We thank A. B. Voitkiv for useful discussions.



\begin{thebibliography}{33}

\bibitem{Review1}
D. B. Milo{\v{s}}evi{\'{c}}, G. G. Paulus, D. Bauer, and W. Becker,
Above-threshold ionization by few-cycle pulses,
J. Phys. B: At. Mol. Opt. Phys. {\bf 39}, R203--R262 (2006);
C. Winterfeldt, C. Spielmann, and G. Gerber,
Colloquium: Optimal control of high-harmonic generation,
Rev. Mod. Phys. {\bf 80}, 117 (2008).

\bibitem{atto}
F. Calegari, G. Sansone, S. Stagira, C. Vozzi and M. Nisoli,
Advances in attosecond science,
J. Phys. B: At. Mol. Opt. Phys. {\bf 49}, 062001 (2016).

\bibitem{Review_solids}
S. Ghimire, G. Ndabashimiye, A. D DiChiara, E. Sistrunk, 
M. I. Stockman, P. Agostini, L. F. DiMauro and D. A Reis,
Strong-field and attosecond physics in solids,
J. Phys. B: At. Mol. Opt. Phys. {\bf 47}, 1 (2014)

\bibitem{solids1}
G. Vampa, T. J. Hammond, N. Thir\'e, B. E. Schmidt, F. L\'egar\'e, 
C. R. McDonald, T. Brabec, D. D. Klug, and P.~B. Corkum,
All-Optical Reconstruction of Crystal Band Structure,
Phys. Rev. Lett. {\bf 115}, 193603 (2015).

\bibitem{solids2}
G. Ndabashimiye, S. Ghimire, M. Wu, D. A. Browne, 
K. J. Schafer, M. B. Gaarde, and D. A. Reis, 
Solid-state harmonics beyond the atomic limit,
Nature {\bf 534}, 520 (2016).

\bibitem{Bauer} D. Bauer and K. K. Hansen,
High-harmonic generation in solids with and without topological edge states,
Phys. Rev. Lett. \textbf{120}, 177401 (2018);
H. Dr\"ueke and D. Bauer,
Robustness of topologically sensitive harmonic generation in laser-driven linear chains,
Phys. Rev. A \textbf{99}, 053402 (2019).

\bibitem{graphene}
A. H. Castro Neto, F. Guinea, N. M. R. Peres, K. S. Novoselov, and A. K. Geim, 
The electronic properties of graphene, 
Rev. Mod. Phys. {\bf 81}, 109 (2009).

\bibitem{Hommelhoff} 
T. Higuchi, C. Heide, K. Ullmann, H. B. Weber, and P. Hommelhoff,
Light-field-driven currents in graphene,
Nature \textbf{550}, 224 (2017);
C. Heide, T. Higuchi, H. B. Weber, and P. Hommelhoff,
Coherent Electron Trajectory Control in Graphene,
Phys. Rev. Lett. \textbf{121}, 207401 (2018).

\bibitem{Hommelhoff_NJP} 
C. Heide, T. Boolakee, T. Higuchi, H. B. Weber, and P. Hommelhoff,
Interaction of carrier envelope phase-stable laser pulses with graphene: 
the transition from the weak-field to the strong-field regime,
New J. Phys. \textbf{21}, 045003 (2019).

\bibitem{Review_top}
X.-L. Qi and S.-C. Zhang,
Topological insulators and superconductors,
Rev. Mod. Phys. {\bf 83}, 1057 (2011).

\bibitem{surface}
J. Reimann {\it et al.}, 
Subcycle  observation of lightwave-driven Dirac currents in a topological surface band,
Nature {\bf 562}, 396 (2018).

\bibitem{Klein} M. I. Katsnelson, K. S. Novoselov, and A. K. Geim,
Chiral tunnelling and the Klein paradox in graphene,
Nature Phys. {\bf 2}, 620 (2006).

\bibitem{Casimir}
A. A. Banishev, H. Wen, J. Xu, R. K. Kawakami, G. L. Klimchitskaya, V. M. Mostepanenko, and U. Mohideen,
Measuring the Casimir force gradient from graphene on a SiO$_2$ substrate,
Phys. Rev. B {\bf 87}, 205433 (2013).

\bibitem{supercriticality}
V. M. Pereira, J. Nilsson, and A. H. Castro Neto,
Coulomb Impurity Problem in Graphene,
Phys. Rev. Lett. {\bf 99}, 166802 (2007).

\bibitem{McGady} D. Allor, T. D. Cohen, and D. A. McGady, 
The Schwinger mechanism and graphene, 
Phys. Rev. D {\bf 78}, 096009 (2008).

\bibitem{Lewkowicz} M. Lewkowicz and B. Rosenstein, 
Dynamics of Particle-Hole Pair Creation in Graphene, 
Phys. Rev. Lett. {\bf 102}, 106802 (2009).

\bibitem{Moessner}
B. Dora and R. Moessner, 
Nonlinear electric transport in graphene: quantum quench dynamics and the Schwinger mechanism, 
Phys. Rev. B {\bf 81}, 165431 (2010).

\bibitem{Mostepanenko}
G. L. Klimchitskaya and V. M. Mostepanenko, 
Creation of quasiparticles in graphene by a time-dependent electric field, 
Phys. Rev. D {\bf 87}, 125011 (2013).

\bibitem{Avetissian}
H. K. Avetissian, A. K. Avetissian, G. F. Mkrtchian, and Kh. V. Sedrakian, 
Creation of particle-hole superposition states in graphene at multiphoton resonant excitation by laser radiation, 
Phys. Rev. B {\bf 85}, 115443 (2012).

\bibitem{Fillion-Gourdeau} F. Fillion-Gourdeau and S. MacLean, 
Time-dependent pair creation and the Schwinger mechanism in graphene, 
Phys. Rev. B {\bf 92}, 035401 (2015).

\bibitem{Akal2016} I. Akal, R. Egger, C. M\"uller, and S. Villalba-Ch\'avez,
Low-dimensional approach to pair production in an oscillating electric field: Application to bandgap graphene layers,
Phys. Rev. D \textbf{93}, 116006 (2016).

\bibitem{Akal2019} I. Akal, R. Egger, C. M\"uller, and S. Villalba-Ch\'avez,
Simulating dynamically assisted production of Dirac pairs in gapped graphene monolayers,
Phys. Rev. D \textbf{99}, 016025 (2019).

\bibitem{bandgap}
Various techniques allow to induce a bandgap in graphene, such as epitaxial growth on suitable substrates, elastic strain, or Rashba spin splittings on magnetic substrates; see, e.g.,  
S. Y. Zhou, G.-H. Gweon, A. V. Fedorov, P. N. First, W. A. de Heer, D.-H. Lee, F. Guinea, A. H. Castro Neto, and A. Lanzara, Substrate-induced bandgap opening in epitaxial graphene,
Nat. Mater. {\bf 6}, 770 (2007);
A. Varykhalov, J. S\'anchez-Barriga, A. M. Shikin, C. Biswas, E. Vescovo, A. Rybkin, D. Marchenko, and O.
Rader, Electronic and Magnetic Properties of Quasifreestanding Graphene on Ni,
Phys. Rev. Lett. {\bf 101}, 157601 (2008).

\bibitem{BW}
G. Breit and J. A. Wheeler, 
Collision of Two Light Quanta, 
Phys. Rev. {\bf 46}, 1087 (1934).

\bibitem{Ritus} V. I. Ritus, 
Quantum effects of the interaction of elementary particles with an intense electromagnetic field,
J. Sov. Laser Res. \textbf{6}, 497-617 (1985).

\bibitem{other}
For recent theoretical work on the NLBW process, 
see the following papers and references cited therein:
A.~Di~Piazza,
Nonlinear Breit-Wheeler pair production in a tightly focused laser beam,
Phys. Rev. Lett. {\bf 117}, 213201 (2016);
S. Meuren, K. Z. Hatsagortsyan, C. H. Keitel, and A.~Di~Piazza,
High-Energy Recollision Processes of Laser-Generated Electron-Positron Pairs,
Phys. Rev. Lett. {\bf 114}, 143201 (2015);
M. J. A. Jansen, J.~Z. Kaminski, K. Krajewska and C. M\"uller,
Strong-field Breit-Wheeler pair production in short laser pulses: Relevance of spin effects,
Phys. Rev. D {\bf 94}, 013010 (2016);
Q.~Z.~Lv, S.~Dong, Y.~T.~Li, Z.~M.~Sheng, Q.~Su, and R.~Grobe,
Role of the spatial inhomogeneity on the laser-induced vacuum decay,
Phys. Rev. A {\bf 97}, 022515 (2018);
A. I. Titov, H. Takabe, and B. K\"ampfer,
Nonlinear Breit-Wheeler process in short laser double pulses,
Phys. Rev. D \textbf{98}, 036022 (2018).

\bibitem{SLAC} D. L. Burke {\it et al.}, 
Positron Production in Multiphoton Light-by-Light Scattering,
Phys. Rev. Lett. {\bf 79}, 1626 (1997).

\bibitem{Pike}  
O.~J.~Pike, F.~Mackenroth, E.~G.~Hill, and S.~J.~Rose,
A photon–photon collider in a vacuum hohlraum,
Nat. Photonics {\bf 8}, 434 (2014);
B.~King, H.~Gies, and A.~Di~Piazza,
Pair production in a plane wave by thermal background photons,
Phys.\ Rev.\ D\ {\bf 86}, 125007 (2012).

\bibitem{Lobet}
X.~Ribeyre, E.~d'Humieres, O.~Jansen, S.~Jequier, V.~T.~Tikhonchuk, and M.~Lobet,
Pair creation in collision of $\gamma$-ray beams produced with high-intensity lasers,
Phys. Rev. E {\bf 93}, 013201 (2016).

\bibitem{Drebot}
I.~Drebot, D.~Micieli, E.~Milotti, V.~Petrillo, E.~Tassi, and L.~Serafini,
Matter from light-light scattering via Breit-Wheeler events produced by two interacting Compton sources,
Phys. Rev. Accel. Beams {\bf 20}, 043402 (2017).

\bibitem{Yu}
J. Q. Yu, H. Y. Lu, T. Takahashi, R. H. Hu, Z. Gong, W. J. Ma, Y. S. Huang, C. E. Chen, and X. Q. Yan,
Creation of Electron-Positron Pairs in Photon-Photon Collisions Driven by 10-PW Laser Pulses,
Phys. Rev. Lett. {\bf 122}, 014802 (2019).

\bibitem{Wallace}
P.~R.~Wallace, The band theory of graphite, 
Phys. Rev.  {\bf 71}, 622   (1947).
  
\bibitem{Gusynin}
S. G. Sharapov, V. P. Gusynin, and H. Beck, 
Magnetic oscillations in planar systems with Dirac-like spectrum of quasiparticle excitations, 
Phys. Rev. B {\bf 69}, 075104 (2004);
V. P. Gusynin and S. G. Sharapov, 
Transport of Dirac quasiparticles in graphene: Hall and optical conductivities, 
Phys. Rev. B {\bf 73}, 245411 (2006).

\bibitem{interaction} 
In the given Lagrangian we do not include the quasiparticle interaction, and, therefore, omit the Coulomb part.

\bibitem{B-field}
The magnetic field component is perpendicular to the graphene sheet. In a classical picture, the particle dynamics is thus restricted to the graphene plane since the Lorentz force $\sim {\bf p}\times{\bf B}$ lies within the plane.

\bibitem{Fradkin}
E. S. Fradkin, D. M. Gitman and S. M. Shvartsman, {\it Quantum Electrodynamics with Unstable Vacuum}, (Springer-Verlag Berlin Heidelberg, 1961).

\bibitem{Varro} For related problems, see
S. Varr\'o, 
New exact solutions of the Dirac equation of a charged particle interacting with an electromagnetic plane wave in a medium,
Laser Phys. Lett. {\bf 10}, 095301 (2013);
M. Oliva-Leyva and G. G. Naumis,
Sound waves induce Volkov-like states, band structure and collimation effect in graphene,
J. Phys.: Condens. Matter {\bf 28}, 025301 (2016).

\bibitem{balance}
Note that the energy-momentum balance for (N)LBW pair production in graphene can be fulfilled by the quasiparticles alone, without participation of the ionic lattice to absorb recoil momentum like it is required in photo-induced Landau-Zener transitions.

\bibitem{corrections} 
Throughout this paper, we consider the rates to leading order in the finestructure constant $\alpha$.
We point out that the Breit-Wheeler rate in QED \cite{BW} attains a very small finite threshold value when next-to-leading order corrections in $\alpha$ due to final-state Coulomb interactions are taken into account; see, e.g., R. J. Gould, Born-approximation and radiative corrections to pair production in photon-photon collisions, Astrophys. J. {\bf 337}, 950 (1989).

\bibitem{threshold1}
E. P. Wigner,
On the Behavior of Cross Sections Near Threshold,
Phys. Rev. {\bf 73}, 1002 (1948);
L. D. Landau and E. M. Lifshitz, {\it Quantum Mechanics} (Pergamon, 1965); Sec. 146.

\bibitem{threshold2}
H. R. Sadeghpour, J. L. Bohn, M. J. Cavagnero, B. D. Esryk, 
I. I. Fabrikant, J. H. Macek and A. R. P. Rau,
Collisions near threshold in atomic and molecular physics,
J.~Phys.~B: At. Mol. Opt. Phys. {\bf 33}, R93-R140 (2000). 

\bibitem{Akhiezer}
A. I. Akhiezer and V. B. Berestetskii, {\it Quantum Electrodynamics}, 
2nd ed. (Wiley, New York, 1965).

\bibitem{terahertz}
D. J. Cook and R. M. Hochstrasser, 
Intense terahertz pulses by four-wave rectification in air,
Opt. Lett. {\bf 25}, 1210 (2000);
T. Bartel, P. Gaal, K. Reimann, M. Woerner, and T. Elsaesser, 
Generation of single-cycle THz transients with high electric-field amplitudes,
Opt. Lett. {\bf 30}, 2805 (2005).

\end{thebibliography}
\end{document}